\documentclass[floatfix,preprint,onecolumn]{revtex4}%
\usepackage{graphicx}
\usepackage{amsmath}
\usepackage{amsfonts}
\usepackage{amssymb}%
\begin{document}
\title{Physical picture for fractures in stratified materials: viscoelastic effects
in large cracks}
\author{Ko Okumura}
\email{okumura@phys.ocha.ac.jp}
\affiliation{Physique de la Mati\`{e}re Condens\'{e}e, Coll\`{e}ge de France, 11 place
Marcelin-Berthelot, 75231 Paris cedex 05, France\linebreak\ and\linebreak
Department of Physics, Graduate School of Humanities and Sciences, Ochanomizu
University, 2--1--1, Otsuka, Bunkyo-ku, Tokyo 112-8610, Japan}
\date{\today}

\begin{abstract}
We present an intuitive physical picture for fractures in nacre-type
stratified materials via scaling arguments: strain distributions around a
fracture are rather different depending on directions and size of fractures.
We thus observe that viscoelastic effects are important for parallel
fractures. This effect are taken into account via the simplest viscoelastic
model for weakly cross-linked polymer. Within a certain limit, we find a
trumpet crack shape similar to that in certain polymers and make predictions
for the fracture energy of certain stratified materials in this mode.

\end{abstract}
\maketitle

\section{Introduction}

Strong structures created in nature through a long history of natural
selection sometimes take advantage of composite structures such as laminar
structures (tooth, timber etc.) where soft and hard layers are intertwined.
Biomimetics aims at exploiting such ideas and some industrial materials (from
blockboard to high-tech tires) have utilized the concept of stacking lamellae
to enhance their toughness \cite{cmp1,cmp2,cmp3,cmp4,cmp5}. In a series of
papers \cite{v1,v2,v3}, we have studied another example of such soft-hard
laminar structure on nanoscales, \textit{nacre}
\cite{nacre1,nacre2,nacre3,nacre4}, to present physical pictures for this substance.

\begin{figure}[ptbh]
\begin{center}
\includegraphics[scale=0.5]{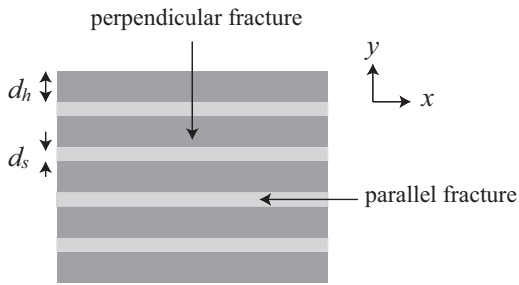}
\end{center}
\caption{Narce-type structure of materials: hard layers (the elastic modulus
$E_{h}$) has a typical thickness $d_{h}$, which are glued together by soft
layers (the elastic modulus $E_{s}$) of a typical thickness $d_{s}$. An
important conditions for these parameters are $\varepsilon_{0}=E_{s}%
/E_{h},\varepsilon_{d}=d_{s}/d_{h}$ $\ll1$ with $\varepsilon=\varepsilon
_{0}/\varepsilon_{d}\ll1$. The cracks in the $y-z$ plane and the $x-z$ plane
are called a perpendicular and parallel fractures, respectively (the $z$-axis
is perpendicular to this page).}%
\label{f1}%
\end{figure}

The structure and notations are illustrated in Fig. \ref{f1}. A typical
thickness $d_{s}$ of the soft layer made from a protein, \textit{conchiolin},
is dozens of nanometer, which is much smaller than a micrometric thickness
$d_{s}$ of the hard layer composed of \textit{aragonite}. An elastic modulus
$E_{s}$ of the soft layer is also much smaller than $E_{h}$ of the hard layer
where the latter is roughly given by 50 GPa. We define two small quantities:%
\begin{align}
\varepsilon_{d}  &  =d_{s}/d_{h}\\
\varepsilon_{0}  &  =E_{s}/E_{h}%
\end{align}
An important property of nacre is that $\varepsilon$ defined by
\begin{equation}
\varepsilon\equiv\varepsilon_{0}/\varepsilon_{d}%
\end{equation}
is very small. We will work in the limit of $\varepsilon_{0},\varepsilon
_{d},\varepsilon\ll1$, and also in the continuum limit where relevant length
scales are much larger than the layer spacing $d$:%
\begin{equation}
d=d_{s}+d_{h}.
\end{equation}
We emphasize here that under these conditions our theory is applicable to a
general layered structure other than nacre. Physically, $\varepsilon_{d}%
$\textit{ has to be small to make a relatively ridged material (as a whole),
while }$\varepsilon$ \textit{(and thus }$\varepsilon_{0}$\textit{) has to be
small to make the stress concentration around tip small to enhance the
toughness}; these conditions seem to have been indeed selected in nacre by nature.

We have been studied both perpendicular and parallel fractures (see Fig.
\ref{f1}) within the linear elastic fracture mechanics (LEFM \cite{Anderson})
under the plain strain condition (the sample is thick in the $z$-direction)
\cite{v1,v2,v3}. One of our aims here is to include a linear viscoelastic
effect into this system.

\section{Novel views on previous results}

The proposed elastic energy per unit volume of the stratified system is given
by \cite{v2}%
\begin{equation}
f=E_{h}e_{xx}^{2}+E_{0}e_{yy}^{2}+E_{0}e_{xy}^{2}+E_{0}e_{xx}e_{yy}%
\end{equation}
where%
\begin{equation}
E_{0}=\varepsilon E_{h},
\end{equation}
We have introduced the strain field
\begin{equation}
e_{ij}=\frac{1}{2}\left(  \frac{\partial u_{i}}{\partial x_{j}}+\frac{\partial
u_{j}}{\partial x_{i}}\right)  \label{e}%
\end{equation}
for the deformation field $u_{i}$ where $(x_{1},x_{2},x_{3})=(x,y,z)$. This
energy reflects, via $\sigma_{ij}=\partial f/\partial e_{ij},$ intuitively
self-evident properties that (1) this system is hard for a tensile stress in
the $x$-direction ($\sigma_{xx}\simeq E_{h}e_{xx}$) due to hard layers, (2) it
is soft for a tensile stress in the $y$-direction ($\sigma_{yy}\simeq
E_{0}e_{yy}$) due to soft layers, and (3) it is soft for a shear stress
($\sigma_{xy}\simeq E_{0}e_{xy}$) again due to soft layers. Here, we notice
that the effectiveness of soft layers is diminished ($E_{s}\rightarrow E_{0}$)
by the thin thickness of the soft layers.

For parallel fractures, we have shown that a usually negligible bending term
becomes important \cite{v3}, which shall be confirmed below from a different
viewpoint. The bending term is given by
\begin{equation}
f_{B}=K_{B}\left(  \frac{\partial^{2}u_{y}}{\partial x^{2}}\right)  ^{2}%
\end{equation}
where%
\begin{equation}
K_{B}=E_{h}d^{2}\equiv E_{0}l^{2}.
\end{equation}
Here we have introduced a length scale,
\begin{equation}
l=d/\sqrt{\varepsilon}, \label{l}%
\end{equation}
which is much longer than the layer period $d$. This bending term becomes
important also in certain liquid crystals \cite{lenticular,LQ}. However, there
is a significant difference from the present situation; in liquid crystals $l$
is replaced by a much smaller length of the order of atomic scales, which
results in a strain distribution quite different from nacre (see below).

\subsection{Scaling structure of elastic energy}

An important difference from the conventional isotropic elastic theory is
that, in an anisotropic system such as nacre, there exist two distinct length
scales $X$ and $Y$ . On the contrary, in isotropic systems at equilibrium when
the elastic body is deformed over a range of $R$ in the $x$-direction, the
deformation relaxes out at the same distance $R$ also in the $y$-direction;
this is because the deformation field of an isotropic system satisfies an
elliptic differential equation with a unique length scale (under the plain
strain condition). In other words, the deformation in nacre does not satisfy
the Laplace equation. Note here that, in the context of a fracture problem,
$X$ corresponds to a length of parallel fracture while $Y$ to that of
perpendicular fracture (see Fig. \ref{f2} below).

Thus, a local energy (per unit volume) of nacre is dimensionally expressed as%
\begin{align}
f  &  \simeq E_{h}\left(  \frac{u_{x}}{X}\right)  ^{2}+E_{0}\left(
\frac{u_{y}}{Y}\right)  ^{2}\nonumber\\
&  +E_{0}\left(  \frac{u_{x}}{Y}+\frac{u_{y}}{X}\right)  ^{2}+E_{0}\frac
{u_{x}}{X}\cdot\frac{u_{y}}{Y}+K_{B}\left(  \frac{u_{y}}{X^{2}}\right)  ^{2}
\label{fsc}%
\end{align}
At equilibrium (statics), $f$ is locally minimized for both variables $u_{x}$
and $u_{y}$; we have two equations for $u_{x}$ and $u_{y}$, i.e. $\partial
f/\partial u_{x}=0$ and $\partial f/\partial u_{y}=0$. Seeking the relevant
solution by requiring $(u_{x},u_{y})\neq(0,0)$, we have an equation for $X$
and $Y$. This equation has two solution specifying a relation between $X$ and
$Y$; one corresponds to perpendicular fractures while the other to parallel
fractures. We should note that here and hereafter a limitation of this type of
dimensional expressions: relative signs (or phases of complex numbers)
sometimes can not be specified.

\subsubsection{Parallel fractures}

The solution for parallel fractures announces a relation, $X^{2}\simeq
Y^{2}\left(  1+l^{2}/X^{2}\right)  $, which implies two regimes for this
problem: $X\gg l$ and $X\ll l$.

For large fractures ($X\gg l$), we have an isotropic relation,
\begin{equation}
X\simeq Y. \label{XY1}%
\end{equation}
On the other hand, with an aide of a relation $\partial f/\partial u_{x}=0$
together with Eq. (\ref{XY1}), we obtain an anisotropic deformation field,
\begin{equation}
u_{x}\simeq\varepsilon u_{y}; \label{uxuy1}%
\end{equation}
we find $u_{x}\ll u_{y}$, which is appropriate for parallel fractures. Using
Eqs. (\ref{XY1}) and (\ref{uxuy1}), we can estimate magnitudes of each term in
Eq. (\ref{fsc}). Keeping terms only at the leading order in $\varepsilon$, we
arrive at%
\begin{equation}
f\simeq E_{0}\left(  \frac{u}{R}\right)  ^{2}%
\end{equation}
where we denote scales of $X\left(  \simeq Y\right)  $ and $u_{y}$ as $R$ and
$u$, respectively. This corresponds to a more precise form $f\simeq
E_{0}\left(  \frac{\partial u_{y}}{\partial y}\right)  ^{2}+E_{0}\left(
\frac{\partial u_{y}}{\partial x}\right)  ^{2}$ shown in \cite{v2}.

For small fractures ($X\ll l$), we have instead
\begin{equation}
X^{2}\simeq Yl, \label{x2yl}%
\end{equation}
which implies $X\gg Y$. The deformation field satisfies
\begin{equation}
u_{x}\simeq\frac{l}{X}\varepsilon u_{y}%
\end{equation}
Here, we note an important condition for a continuum theory: $X,Y>d$. This
condition, together with Eq. (\ref{l}), results in
\begin{equation}
\varepsilon l^{2}/X^{2},\varepsilon l^{2}/Y^{2}<1
\end{equation}
which implies $\varepsilon l/X,\varepsilon l/Y<\sqrt{\varepsilon}\ll1$ (note
also that this does not necessarily imply $l<X,Y$; e.g. $X\gtrsim
\sqrt{\varepsilon}l$ satisfies both $l>X$ and $\varepsilon l/X<1$). Thus, we
again find $u_{x}\ll u_{y}$. In this way, we arrive at%
\begin{equation}
f\simeq E_{0}\left(  \frac{u}{Y}\right)  ^{2}\simeq K_{B}\left(  \frac
{u}{X^{2}}\right)  ^{2}%
\end{equation}
which corresponds to $f\simeq E_{0}\left(  \frac{\partial u_{y}}{\partial
y}\right)  ^{2}+K_{B}\left(  \frac{\partial^{2}u_{y}}{\partial x^{2}}\right)
^{2}$ shown in \cite{v3}.

\subsubsection{Perpendicular fractures}

The solution for perpendicular fractures announces an anisotropic relations,
\begin{align}
Y  &  \simeq\sqrt{\varepsilon}X\,\\
u_{y}  &  \simeq\sqrt{\varepsilon}u_{x}%
\end{align}
for which $u_{x}\gg u_{y}$. These relations lead to an energy,%
\begin{equation}
f\simeq E_{h}\left(  \frac{u}{X}\right)  ^{2}\simeq E_{0}\left(  \frac{u}%
{Y}\right)  ^{2}.
\end{equation}
Here, $u$ denotes not $u_{y}$ but $u_{x}$. This energy corresponds to $f\simeq
E_{h}\left(  \frac{\partial u_{x}}{\partial x}\right)  ^{2}+E_{0}\left(
\frac{\partial u_{x}}{\partial y}\right)  ^{2}$ shown in \cite{v2}.

\subsection{Scaling views on fracture mechanics}

\begin{figure}[ptbh]
\begin{center}
\includegraphics[scale=0.5]{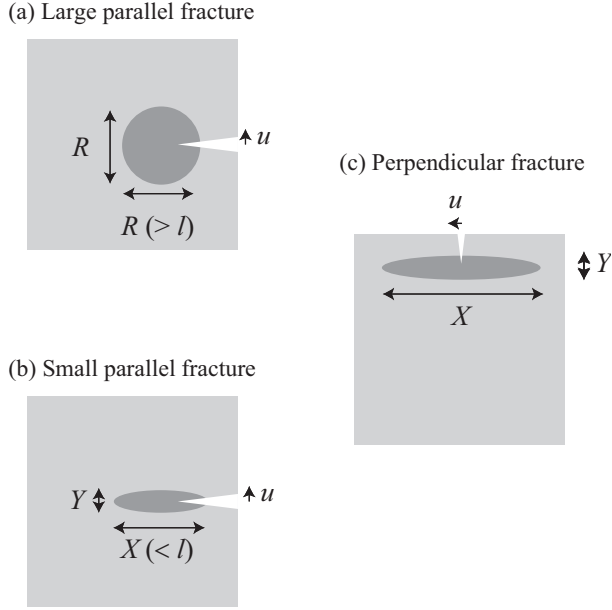}
\end{center}
\caption{Physical images of strain distribution accompanying a fracture. Dark
areas indicate the region where the strain field is significant. (a) Large
parallel fracture. There is only one length scale $R$ ($\gg l$) for the
distribution, where the crack length is also of the order of $R$. (b) Small
parallel fracture. The strain distribution becomes rather anisotropic: the
length scale $X$ ($\ll l$) for $x$-direction is larger than $Y$, where the
crack size is of the order of $X$. (c) Perpendicular fracture. An anisotropic
strain distribution with $X\gg Y$, where the crack size is of the order of
$Y$.}%
\label{f2}%
\end{figure}

We consider a fracture in nacre-type materials. Strain distribution for three
types of fractures are schematically given in Fig. \ref{f2}. We should note
that these figures are \textit{at the scaling level }and thus they are rather
rough images; some examples of comparison with exact results are given in
Appendix. We will develop below scaling arguments in order to reproduce our
previous results but only dimensionally. Similar arguments in a different
context was first presented in \cite{lenticular}. 

\subsubsection{Parallel fractures}

For \textit{large cracks} ($X\gg l$), the strain distribution is isotropic
(Fig. \ref{f2}a) and, thus, the potential energy per unit crack-front length
(in $z$-direction) is dimensionally given by%
\begin{equation}
F\simeq E_{0}\left(  \frac{u}{R}\right)  ^{2}R^{2}-\sigma uR+2G_{\parallel}R
\end{equation}
where the first two terms describes the elastic potential energy and the last
term an energy loss due to a creation of new surfaces in this system:
$G_{\parallel}$ is the fracture energy (energy required to create a unit
area). We minimize this energy with respect to $u$ to find a Hooke's law%
\begin{equation}
\sigma\simeq E_{0}u/R \label{Hooke}%
\end{equation}
which leads to an optimized energy value,%
\begin{equation}
F_{m}\simeq-\sigma^{2}R^{2}/E_{0}+2G_{\parallel}R,
\end{equation}
which is quadratic in $R$; the maximum is given at $R=R^{\ast}$:%
\begin{equation}
\sigma\simeq\sqrt{G_{\parallel}E_{0}/R^{\ast}}. \label{sigma1}%
\end{equation}
When $R<R^{\ast}$, $F_{m}$ decreases with decrease in $R$; a fracture with the
size smaller than $R^{\ast}$ tends to close to lower the energy. On the
contrary, when $R>R^{\ast}$, $F_{m}$ decreases with increase in $R$; a
fracture with the size larger than $R^{\ast}$ tends to expand. Thus, Eq.
(\ref{sigma1}) corresponds to a critical failure stress. From Eqs.
(\ref{Hooke}) and (\ref{sigma1}), we have
\begin{equation}
u\simeq\sqrt{G_{\parallel}R^{\ast}/E_{0}} \label{u1}%
\end{equation}
Eqs. (\ref{sigma1}) and (\ref{u1}) give scaling structures for the stress and
deformation fields obtained more precisely in \cite{v2}: the stress scales as
$R^{-1/2}$ and the deformation is parabolic$~\left(  \sim\sqrt{R}\right)  $ as
in the conventional LEFM. They also give a scaling relation%
\begin{equation}
\sigma u\simeq G_{\parallel}, \label{su1}%
\end{equation}
which announces that the product $\sigma u$ (at critical state, $R\simeq
R^{\ast}$) gives the fracture energy (see Discussion). The fracture energy
$G_{\parallel}$ might be associated with the separation energy of the soft
layer and we previously concluded that there was no significant enhancement in
this mode. Due to the ensuing analysis we can afford a more precise
understanding on this.

In conventional LEFM, the stress intensity factor $K$ scales as $\sqrt
{\mathcal{G}E_{0}}$ in the present context where $\mathcal{G}$ is \emph{the
energy release rate} and \textit{its critical value} is the fracture energy
$\left(  G_{\parallel}\right)  $. Eq. (\ref{sigma1}) suggests that $K$ scales
not as $\sqrt{\mathcal{G}E_{0}}$ but as its critical value$\sqrt{G_{\parallel
}E_{0}}$; this is logical because in Eqs. (\ref{sigma1})-(\ref{su1}) we are
always at a critical of failure ($R\simeq R^{\ast}$).

For \textit{small cracks} ($X\ll l$), the strain distribution is anisotropic
(Fig. \ref{f2}b). We note here that this distribution is quite different from
certain cases of liquid crystals \cite{lenticular}; there strain is
distributed more widely in $y$-direction than in $x$-direction. The potential
of our system is given by%
\begin{equation}
F\simeq K_{B}\left(  \frac{u}{X^{2}}\right)  ^{2}XY-\sigma uX+2G_{\parallel}X.
\end{equation}
We minimize this energy with respect to $u$ to find a Hooke's law%
\begin{equation}
\sigma\simeq E_{0}u/Y
\end{equation}
which corresponds to $\sigma_{yy}\simeq E_{0}\partial u_{y}/\partial y$. Note
here that the dominant component of stress field tensor is $\sigma_{yy}$
$\left(  \text{e.g. }\sigma_{xy}\simeq E_{0}\partial u_{y}/\partial x\simeq
E_{0}(u/X)\ll\sigma_{yy}\text{ due to Eq. (\ref{x2yl})}\right)  $. The energy
optimized for $u$ is here given by
\begin{equation}
F_{m}\simeq-\sigma^{2}X^{3}l/K_{B}+2G_{\parallel}X.
\end{equation}
$F_{m}$ takes its maximum value at $X\simeq X^{\ast}\simeq\sqrt{Y^{\ast}l}$,
where%
\begin{align}
\sigma &  \simeq\sqrt{\frac{K_{B}G_{\parallel}}{l}}\frac{1}{X^{\ast}}%
\simeq\sqrt{\frac{G_{\parallel}E_{0}}{Y^{\ast}}},\label{sigma2}\\
u  &  \simeq\sqrt{\frac{lG_{\parallel}}{K_{B}}}X^{\ast}\simeq\sqrt
{\frac{G_{\parallel}Y^{\ast}}{E_{0}}}, \label{u2}%
\end{align}
Eqs. (\ref{sigma2}) and (\ref{u2}) give scaling structures for the stress and
deformation fields as well as a scaling relation $\sigma u\simeq G_{\parallel
}$. We emphasize here that the stress thus obtained (which is proportional to
strain) is indeed anisotropic; $\sigma\sim1/X\sim1/\sqrt{Yl}$. The stress and
deformation fields obtained in \cite{v3} are also consistent with these
scaling structures; we can check that, dimensionally, $\sigma(x,y)$ and
$u(x,y)$ derived in \cite{v3} indeed reduce to Eqs. (\ref{sigma2}) and
(\ref{u2}) by setting $x^{2}\simeq yl$ (because we are always in the regime
specified by Eq. (\ref{x2yl}) at the scaling level).

\subsubsection{Perpendicular fractures}

The strain distribution is anisotropic (Fig. \ref{f2}c) and, thus, the
potential is given by%
\begin{equation}
F\simeq E_{h}\left(  \frac{u}{X}\right)  ^{2}XY-\sigma uY+2G_{\perp}Y,
\end{equation}
where $G_{\perp}$ is the fracture energy. We minimize this energy with respect
to $u$ to find a Hooke's law%
\begin{equation}
\sigma\simeq E_{h}u/X
\end{equation}
which corresponds to $\sigma_{xx}\simeq E_{h}\partial u_{x}/\partial x$ (the
dominant stress tensor component in this case is $\sigma_{xx}$). The energy
optimized for $u$ is here given by
\begin{equation}
F_{m}\simeq-\frac{\sigma^{2}Y^{2}}{\sqrt{\varepsilon}E_{h}}+2G_{\perp}Y.
\end{equation}
$F_{m}$ takes its maximum value at $Y\simeq Y^{\ast}\simeq\sqrt{\varepsilon
}X^{\ast}$, where%
\begin{align}
\sigma &  \simeq\sqrt{\frac{G_{\perp}E_{h}}{Y^{\ast}/\sqrt{\varepsilon}}%
},\label{38}\\
u  &  \simeq\sqrt{\frac{G_{\perp}}{E_{h}}\left(  Y^{\ast}/\sqrt{\varepsilon
}\right)  }, \label{39}%
\end{align}
Scaling structures for the stress and deformation fields in Eqs. (\ref{38})
and (\ref{39}) are in accord with results in \cite{v2} (the stress or strain
thus obtained is indeed anisotropic). In this case, we can estimate the
fracture energy $\sigma u\simeq G_{\perp}$ via the maximum stress for a
continuum stress $\left(  \sigma_{F}\simeq\sqrt{\sqrt{\varepsilon}G_{\perp
}E_{h}/d}\right)  $ balancing with the yield stress of the pure aragonite
$\left(  \sigma_{YS}\simeq\sqrt{E_{h}\gamma_{h}/a_{h}}\right)  $: $G_{\perp
}\simeq\lambda_{\perp}\gamma_{h}$ where $\lambda_{\perp}=d/\left(
\sqrt{\varepsilon}a_{h}\right)  \gg1$; the fracture energy is enhanced from
that of the pure aragonite $\gamma_{h}$ because $a_{h}\left(  \ll d\right)  $
is a microscopic size of defects in a sample.

\section{Inclusion of viscoelastic effect}

From the above scaling arguments, we have seen that $u_{y}$ is a dominant
component of the deformation field for parallel fractures while $u_{x}$ is
important for perpendicular fractures. Since $u_{y}$ essentially corresponds
to the deformation in soft layers while $u_{x}$ to that in hard layers,
viscoelastic effects associated with extremely soft layers becomes important
only for parallel fractures; we shall include these effects into our theory.

As a model for viscoelasticity, we employ the simplest model (also utilized in
\cite{trumpet}):%
\begin{equation}
\mu(\omega)=\mu_{0}+\left(  \mu_{\infty}-\mu_{0}\right)  \frac{i\omega\tau
}{1+i\omega\tau} \label{trumpet}%
\end{equation}
This is the minimal model to include the following essential properties of
weakly cross-linked network: (1) its slow motions are governed by a
weak-modulus $\mu_{0}$ associated with weak cross-links, (2) its fast motions
involve a strong-modulus $\mu_{\infty}$ ($\gg$ $\mu_{0}$) originating from
entanglements.\begin{figure}[ptbh]
\begin{center}
\includegraphics[scale=0.5]{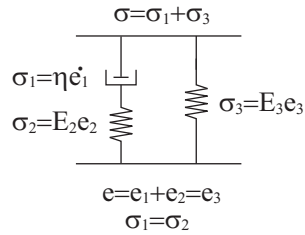}
\end{center}
\caption{Zener model with two springs and a dashpot.}%
\label{f3}%
\end{figure}

This model is a special case of the Zener model (see Fig. \ref{f3}),
\cite{Hui,Oudet} for which stress $\sigma$ and strain $e$ satisfy the
relation,%
\begin{equation}
\sigma+\frac{\eta}{E_{2}}\dot{\sigma}=E_{3}e+\eta\frac{E_{2}+E_{3}}{E_{2}}%
\dot{e} \label{Zener}%
\end{equation}
with identification, $E_{3}=\mu_{0}$, $E_{2}=\mu_{\infty}-\mu_{0},$ and
$\tau=\eta/E_{2}$.

The previous purely elastic treatment should correspond to the zero frequency
limit or the static limit of a viscoelastic model. For parallel fractures,
($y,y$) component of stress ($\sigma$) is important, for which we write
$\sigma=\sigma_{s}\simeq E_{s}e_{s}$ where $\sigma_{s}$ and $e_{s}$ denote
stress and strain of soft layers. Thus, it is natural to require that soft
layers obey the trumpet relation $\sigma_{s}+\tau\dot{\sigma}_{s}=\mu_{0}%
e_{s}+\mu_{\infty}\tau\dot{e}_{s}$ with $\mu_{0}=E_{s}$, because then the
static limit reduces to the desired form. Noting the relations between
coarse-grained stress and strain of nacre ($\sigma$ and $e$) and more local
stress and strain of soft layers ($\sigma_{s}$ and $e_{s}$), i.e.
$\sigma=\sigma_{s}$ and $e_{s}d_{s}=ed$, we deduce a viscoelastic model for
nacre:%
\begin{equation}
\sigma+\tau\dot{\sigma}=E_{0}e+E_{\infty}\tau\dot{e}%
\end{equation}
where%
\begin{align}
E_{\infty}  &  =\lambda E_{0}=\lambda E_{s}/\varepsilon_{d}\\
\tau &  =\frac{\eta}{\mu_{\infty}-\mu_{0}}\\
\lambda &  =\mu_{\infty}/\mu_{0}%
\end{align}
We note again that $\mu_{0}=E_{s}$ is associated with weak cross-links in soft
layers, while $\mu_{\infty}$ with entanglement in soft layers.

Putting $\sigma=\sigma_{m}e^{i\omega t}$ and $e=e_{m}e^{i\omega t}$, with the
definition of the viscoelastic modulus, $\sigma_{m}=\mu(\omega)e_{m}$, we
obtain Eq. (\ref{trumpet}) with $\mu_{\infty}$ and $\mu_{0}$ replaced by
$E_{\infty}$ and $E_{0}$, respectively. This model can be divided into three
regimes:
\begin{equation}%
\begin{array}
[c]{lll}%
\text{I. }\omega\tau\ll1/\lambda\text{:} & \mu(\omega)\simeq E_{0} &
\text{(soft solid)}\\
\text{II. }1/\lambda\ll\omega\tau\ll1\text{:} & \mu(\omega)\simeq i\omega\eta
& \text{ (liquid)}\\
\text{III. }1\ll\omega\tau\text{:} & \text{ }\mu(\omega)\simeq E_{\infty} &
\text{(hard solid)}%
\end{array}
\text{ } \label{trumpetIII}%
\end{equation}

\section{Crack shape and fracture energy}

In the following we consider only large cracks: $l\gg X$. The opposite limit
is examined in a separate article. \cite{EPL} In the case of such large
cracks, we are always in the isotropic regime where $X\simeq Y(\equiv R)$ (see
also Fig. \ref{f2}a). We imagine a crack propagating at a constant speed $V$.
For simplicity, we treat only certain large velocities: $l\ll V\tau$; although
it is very difficult to estimate $\tau$ for the protein, this corresponds to
$\tau\gg10^{-2}$ s for $l\simeq100$ $\mu$m and $V\simeq1$ cm/s.

An important observation here is the scaling identification of a distance $r$
from the tip and the frequency $\omega$ via a stationary crack-propagation
speed $V$:
\begin{equation}
r\simeq V/\omega; \label{rV}%
\end{equation}
small distances correspond to high frequencies while long distances to low
frequencies --- the farther away from the tip, the more time for relaxation.
Corresponding to Eq. (\ref{trumpetIII}), the fracture is spatially divided
into three regions:%
\begin{equation}%
\begin{array}
[c]{ll}%
\text{I. }\lambda V\tau\ll r\text{:} & \text{soft solid}\\
\text{II. }V\tau\ll r\ll\lambda V\tau\text{:} & \text{liquid}\\
\text{III. }d\ll r\ll V\tau\text{:} & \text{hard solid}%
\end{array}
\text{ }%
\end{equation}

Another important observation is that, in certain model of viscoelasticity,
the scaling relation $\sigma\sim r^{-1/2}$ (where $r$ is a distance from the
fracture tip) still holds as in a purely elastic model. This point shall be
explained in next section.

\begin{figure}[ptbh]
\begin{center}
\includegraphics[scale=0.8]{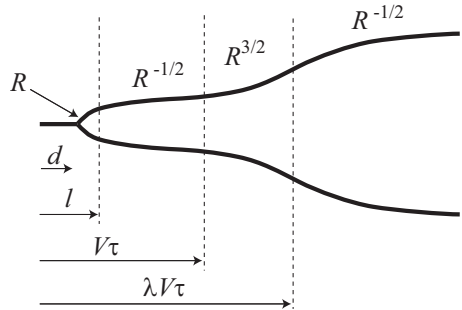}
\end{center}
\caption{Viscoelastic trumpet. The dependence of deformation field $u$ on a
distance from the tip $R$ is indicated in the figure. Very close to the tip
($l\ll R$), the shape \textit{merges to a lenticular form} \cite{v3}.}%
\label{f4}%
\end{figure}

First we consider a large crack size ($L\gg\lambda V\tau$); then, all the
three regions I-III are present. The soft-solid region I corresponds to low
frequencies, and thus, to the static limit; in this region ($r\gg\lambda
V\tau\gg l$), Eqs. (\ref{sigma1})-(\ref{su1}) holds. In the liquid zone II,
the stress field scales as $\sigma\simeq\omega\eta e\simeq\eta Vu/r^{2}$ and
at the same time it should scale as $\sigma\sim r^{-1/2}$ from the above
second observation. Thus, the strain should scale as $u\sim r^{3/2}$ and the
product as $\sigma u\sim r$. The coefficients can be determined by matching at
$r\simeq\lambda V\tau$, i.e. the boundary between I and II: for example, we
have%
\begin{equation}
\sigma u\simeq G_{\parallel}\frac{r}{\lambda V\tau}\text{ \ (liquid zone)}.
\label{suL}%
\end{equation}
When we reach the hard solid region ($r\simeq V\tau$), we find%
\begin{equation}
\sigma u\simeq G_{\parallel}/\lambda\text{ \ (hard-solid zone)} \label{suHS1}%
\end{equation}
When we are in the solid region, as in the same manner (due to the simplifying
condition $l\ll V\tau$) with the derivation of Eqs. (\ref{sigma1}%
)-(\ref{su1}), we can show%
\begin{align}
\sigma &  \simeq\sqrt{G_{0}E_{\infty}/R^{\ast}}\label{sHS}\\
u  &  \simeq\sqrt{G_{0}R^{\ast}/E_{\infty}}\\
\sigma u  &  \simeq G_{0} \label{suHS2}%
\end{align}
Here, we emphasize that $G_{0}$ is associated with the hard solid appearing
near the tip and is different from $G_{\parallel}$ (At the place very close to
the tip ($r\ll l$), Eqs. (\ref{sHS})-(\ref{suHS2}) are no longer valid and
they merge to the lenticular expressions \cite{v3}). From Eqs. (\ref{suHS1})
and (\ref{suHS2}), we find%
\begin{equation}
G_{\parallel}\simeq\lambda G_{0}.
\end{equation}
The overall separation energy $G_{\parallel}$ is enhanced from $G_{0}$
associated with the hard solid. The crack shape resulting from this analysis
is just like a trumpet (see Fig. \ref{f4}), as has been suggested by the name
of the model. The similar form has been predicted \cite{trumpet} and observed
\cite{Ondarcuhu} also in systems (weakly or non cross-linked polymers) rather
different from the present anisotropic system. 

We complete our arguments by considering smaller fractures. When
$(l<)V\tau<L<\lambda V\tau$, only the hard-solid and liquid regions are
present; the soft solid has yet to develop. In this situation, the fracture
energy is given by Eq. (\ref{suL}) at $r=L$: $G_{\parallel}(V)=G_{0}L/(V\tau
)$; the toughness decreases with velocity. When $(l<)L<V\tau$, only the
hard-solid region is developed and the fracture energy is given by $G_{0}$.
Thus, as a function of $V$, the fracture energy starts from a larger plateau
value $\lambda G_{0}$, and then decreases to reach a smaller plateau value
$G_{0}$:%
\begin{equation}
G_{\parallel}(V)=\left\{
\begin{array}
[c]{ll}%
\lambda G_{0} & \text{ for }d/\tau<V<L/\left(  \lambda\tau\right) \\
G_{0}L/(V\tau) & \text{ for }L/\left(  \lambda\tau\right)  <V<L/\tau\\
G_{0} & \text{ for }L/\tau<V
\end{array}
\right.  \text{ }%
\end{equation}
Note here that in a continuum theory we only consider the region $V\tau>d$.
This behavior can be confirmed more precisely from a general formula
\cite{trumpet}:%
\begin{equation}
\frac{G_{\parallel}(V)}{G_{0}}\simeq E_{\infty}\int\frac{d\omega}{\omega
}\operatorname{Im}\left[  \frac{1}{\mu(\omega)}\right]
\end{equation}
which can be analytically calculated for the trumpet model (just as in the
previously known isotropic case \cite{Florent}).

\section{Discussion}

In \cite{trumpet} the following observations are used: (A1) for a certain
range of $V$, a (dimensional) dynamical equation $\rho Dv/Dt=-\nabla\sigma$
for viscoelastic materials can be replaced by $\nabla\sigma=0$, (A2)
accordingly, in certain model of viscoelasticity, the scaling relation
$\sigma\sim r^{-1/2}$ still holds as in a purely elastic model, (A3) similar
to the usual LEFM, we can expect a relation, $\mathcal{G}\simeq K^{2}%
/E_{\infty}$ for certain viscoelastic models (where $K$ and $\mathcal{G}$ the
stress intensity factor and the energy release rate associated with the tip),
and (A4) we can find scaling structure of fracture energy from the product
$\sigma u$, calculated at a large $r$. (Our arguments are, however, slightly
different from the original one in that we did not explicitly used (A3).) We
present possible interpretations for these observations to complement the
original arguments.

Before showing our interpretations, we assume that the rheological relation is
linear in a sense that the stress and strain are governed by%
\begin{equation}
\sigma(t)=\int dt^{\prime}E(t-t^{\prime})\dot{e}(t^{\prime}) \label{LR}%
\end{equation}
In terms of Fourier transform (in time), this reduces to a simple
\textit{proportional relation}:%
\begin{equation}
\sigma_{\omega}=\mu(\omega)e_{\omega}%
\end{equation}
where $\mu(\omega)=i\omega E_{\omega}$; $\mu(\omega)$ has the dimension of a
elastic modulus $E(t)$. Here, we have used the notation $O_{\omega}=\int
dtO(t)e^{-i\omega t}$. We shall use the scaling identification Eq. (\ref{rV})
in the following arguments.

\subsection{Interpretation for $\nabla\sigma=0$}

$\rho Dv/Dt=-\nabla\sigma$ can be dimensionally reexpressed in terms of
Fourier components (in space and time) of stress and deformation as
$\rho\omega^{2}u_{k,\omega}\simeq k\sigma_{k,\omega}$. This equation reduces
to $V^{2}u_{k,\omega}\simeq c^{2}u_{k,\omega}$ for a linear rheology
($\sigma_{k,\omega}\simeq Eku_{k,\omega}$) and for a crack propagation speed
$V$ ($\omega\simeq Vk,$ or Eq. (\ref{rV})). Here, the sound velocity $c$ is
defined as $c=\sqrt{E/\rho}$. Thus, if $V\ll c$, we can neglect $\rho Dv/Dt$
against $\nabla\sigma$. For $E\simeq10^{5}$ N/m$^{2}$ and $\rho\simeq10^{3}$
kg/m$^{3}$, $c\simeq10$ m/s, which is sometimes high compared with typical
crack propagation speeds, say, of cm/s.

\subsection{Interpretation for $\sigma\sim r^{-1/2}$}

In the plane strain condition, there are only three independent stress
components: $\sigma_{xx}$, $\sigma_{yy}$ and $\sigma_{xy}$. For $V\ll c$, we
can start from $\nabla\sigma=0$, which gives only two independent equations:
$\frac{\partial\sigma_{xx}}{\partial x}+\frac{\partial\sigma_{xy}}{\partial
y}=0$ and $\frac{\partial\sigma_{yy}}{\partial y}+\frac{\partial\sigma_{xy}%
}{\partial x}=0$. The third equation is a compatibility equation, which
directly follows from the definition (Eq. (\ref{e})): $\frac{\partial
^{2}e_{xx}}{\partial y^{2}}+\frac{\partial^{2}e_{yy}}{\partial x^{2}}%
=2\frac{\partial^{2}e_{xy}}{\partial x\partial y}$. Again, for a linear
rheology, we have a simple proportional relation of the type $\sigma
_{ij}=E_{ijkl}e_{kl}$ or $e_{ij}=D_{ijkl}\sigma_{kl}$ in Fourier space (in
time only). Then, the three differential equations for Fourier component (in
time) of $\sigma_{ij}$ with respect to variables $x$ and $y$ have the same
spacial scaling structure as in a purely elastic model where $\sigma
(t)=E(t)e(t)$. Thus, the scaling structure, $\sigma\sim r^{-1/2}$ should be
unaltered even in a linear viscoelastic model.

\subsection{Interpretation for $\mathcal{G}\simeq K^{2}/E_{\infty}$ and
scaling formula for fracture toughness $G\simeq\sigma u$}

\begin{figure}[ptbh]
\begin{center}
\includegraphics[scale=0.7]{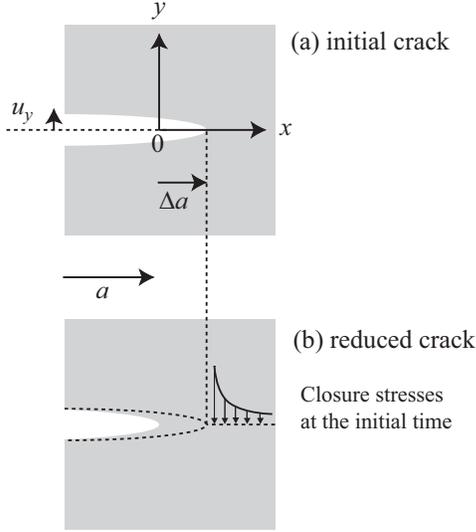}
\end{center}
\caption{Closure of a crack.}%
\label{fa1}%
\end{figure}

To establish $\mathcal{G}\simeq K^{2}/E_{\infty}$ in the viscoelastic model
and to understand why it can not be replaced by $\mathcal{G}\simeq K^{2}%
/E_{0}$, we start from the text book derivation \cite{Anderson}; we imagine
that a crack is closed to a smaller size from the tip by applying a necessary
work, say from a initial size $a+\Delta a$ to a reduced size $a$ (see Fig.
\ref{fa1}).

The energy release rate $\mathcal{G}$ (per unit length of crack front) is
defined by%
\begin{equation}
\mathcal{G}=\lim_{\Delta a\rightarrow0}\frac{\Delta U}{\Delta a} \label{ERR}%
\end{equation}
where the work of crack closure (per unit crack-front length) \ can be
expressed as%
\begin{equation}
\Delta U\simeq\int_{x=0}^{x=\Delta a}2dx\int dy\int_{e}^{0}\left(
-\sigma\right)  de
\end{equation}
Here, the factor 2 counts both the upper and the lower sides of the crack.

This quantity depends on a history of closure process, for example, via Eq.
(\ref{LR}). However, we will finally consider the limit $\Delta a\rightarrow0$
as in Eq. (\ref{ERR}), while the relevant frequency is given by $V/\Delta a$
($\rightarrow\infty$): very high frequency is important. In such a case, the
trumpet model behaves as a solid with a modulus $E_{\infty}$ ($\sigma\simeq
E_{\infty}e$). Namely,
\begin{equation}
\Delta U\simeq\int dx\int dy\sigma^{2}/E_{\infty}\text{.} \label{dU2}%
\end{equation}
As shown before, even in the trumpet model, we expect $\sigma\simeq K/\sqrt
{r}$, if we denote the coefficient (the stress intensity factor) by $K$. Thus,
Eq. (\ref{dU2}) leads to $\Delta U\simeq\Delta aK^{2}/E_{\infty}$; we indeed
recover $\mathcal{G}\simeq K^{2}/E_{\infty}.$ At the critical of failure, this
corresponds to
\begin{equation}
G_{0}\simeq K^{2}/E_{\infty} \label{GK2E}%
\end{equation}
\textit{The critical value of the energy release rate, which is the
conventional fracture energy, is associated here with the hard solid region
near the tip.}

We emphasize here that the overall fracture energy $G_{\parallel}$ should
scale as $\lim_{\Delta a\rightarrow\infty}\frac{\Delta U}{\Delta a}$, which is
\textit{different from the critical value of the energy release rate}. Then,
in contrast with the limit $\Delta a\rightarrow0$, very low frequency becomes
important; we have to use $\sigma\sim E_{0}e$ in this context to find
\begin{equation}
G_{\parallel}\sim K^{2}/E_{0}. \label{GK2E2}%
\end{equation}

The stress intensity factor $K$ in Eqs. (\ref{GK2E}) and (\ref{GK2E2}) are
actually different; these factors are an unknown coefficient for a known
singularity ($\sim\sqrt{r}$) in the above arguments. In fact, from Eqs.
(\ref{sigma1}) and (\ref{sHS}), they are respectively given by $\sqrt
{G_{0}E_{\infty}}$ and $\sqrt{G_{\parallel}E_{0}}$. In other words, Eqs.
(\ref{GK2E}) and (\ref{GK2E2}) can be directly reproduced from Eqs.
(\ref{sigma1}) and (\ref{sHS}).

Noting $\Delta U\simeq\int dx\sigma(x)u(x)$ for linear models ($\sigma\sim
e$), we have an expression of the overall fracture energy $G_{\parallel}$ for
\textit{a macroscopic} $\Delta a$:
\begin{equation}
G_{\parallel}\simeq\frac{1}{\Delta a}\int_{0}^{\Delta a}\sigma(x)u(x)dx
\end{equation}
If the product $\sigma(x)u(x)$ reaches a plateau value at a macroscopic
distance $x_{0}$, we can expect a relation:%
\begin{equation}
G_{\parallel}\simeq\sigma(x_{0})u(x_{0})
\end{equation}
where the left-hand side is evaluated at a macroscopic distance $x_{0}$ from
the tip. This explains why we can use this product to estimate the fracture
toughness, as announced in Sec. II. B (see, for example, Eq. (\ref{su1})).

\section{Conclusion}

In this paper, we present physical pictures for fractures in nacre-type
materials via scaling arguments, which is complementary to our previous more
detailed analysis: strain distributions around a fracture are significantly
different among (small and large) parallel and perpendicular fractures. From
pictures thus obtained, we see that viscoelastic effects may play a role in
parallel fractures and the effects are taken into account via the simplest
model. We limit ourselves to the case of large cracks ($X\gg l$) where
crack-propagation speeds are not too slow ($l\ll V\tau$); beyond this limit,
crack shape and behavior of $G(V)$ seems to be quite different, as discussed
in a separate paper \cite{EPL}. Within this limit, we found that the crack
shape takes the trumpet form as in weakly (or non-) cross-linked polymers
\cite{Ondarcuhu} although they are different in the region very close to the
tip (this difference might be difficult to detect). The overall fracture
energy $G_{\parallel}$ is found to be enhanced from separation energy $G_{0}$
associated with the hard solid region developed near the tip. It is emphasized
that, here, the critical value of the energy release rate corresponds not to
the overall fracture energy but a separation energy associated with the hard solid.

\begin{acknowledgments}
The author (K.O.) is grateful to P.-G. de Gennes for useful discussions. K.O.
also greatly appreciates Elie Rapha\.{e}l and Florent Saulnier for discussions
and for making their recent manuscript \cite{Florent} available prior to
submission. K.O. thanks members of group of P.G.G. at Coll\`{e}ge de France,
including David Qu\'{e}r\'{e}, for a warm hospitality during his third stay in
Paris. This stay is financially supported by Coll\`{e}ge de France.
\end{acknowledgments}

\appendix*

\section{Rough images at the scaling level and exact distributions}

\begin{figure}[ptbh]
\begin{center}
\includegraphics[scale=0.6]{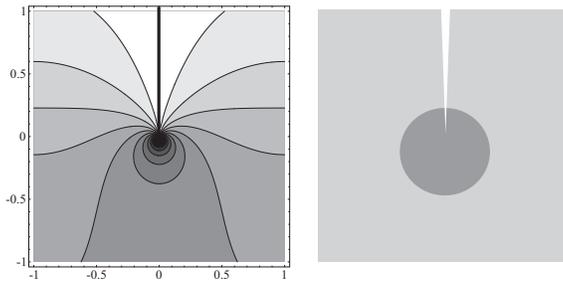}
\end{center}
\caption{Left: contour plot of strain distribution for a crack (size $L$) in
an isotropic elastic sample (the dimension $2L\times2L$) based on an
analytical solution ($\varepsilon=1$). Here, $L$ is the unit length. Right:
corresponding physical image. The distribution is isotropic in a sense that
there is only one length scale $L$.}%
\label{a1}%
\end{figure}

\begin{figure}[ptbhptbh]
\begin{center}
\includegraphics[scale=0.6]{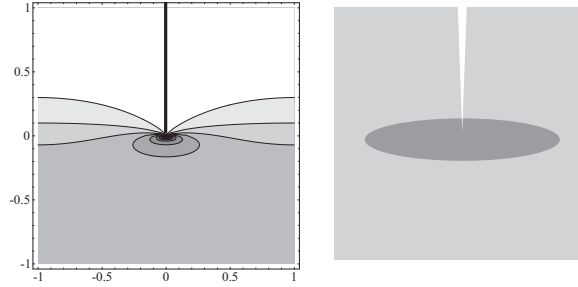}
\end{center}
\caption{Left: contour plot of strain distribution for a perpendicular crack
(size $L$) in an nacre-type elastic sample (the dimension $2L\times2L$) based
on an analytical solution obtained in \cite{v2} ($\varepsilon=0.1$). Here
again, $L$ is the unit length. Right: corresponding physical image. The
distribution is aisotropic in a sense that there is two separate length scales
where the length scale for $x$-direction is $L$ and that for $y$-direction is
$L/\sqrt{\varepsilon}$.}%
\label{a2}%
\end{figure}

In order to clarify the meaning of physical pictures illustrated in Fig.
\ref{f2}, we give examples on comparison of such pictures with an exact
solution available for perpendicular cracks (Figs. \ref{a1} and \ref{a2}). We
employ the following analytical expression \cite{v2}:%
\begin{equation}
\sigma\sim e\sim\operatorname{Re}\left[  \frac{e^{i\pi z/\left(  2L\right)  }%
}{\left(  e^{i\pi z/L}-1\right)  ^{1/2}}\right]
\end{equation}
with $z=x+iy/\sqrt{\varepsilon}$ where $\varepsilon=1$ and $\varepsilon=0.1$
for Figs. \ref{a1} and \ref{a2}, respectively ($L=1$). This expression have
separate length scales for $x$- and $y$-direction: $L$ and $L/\sqrt
{\varepsilon}$, respectively. Fig. \ref{a1} presents an example of the
conventional isotropic sample ($\varepsilon=1$) while Fig. \ref{a2} an example
of a perpendicular crack in a nacre-type material ($\varepsilon=0.1$%
).

In most experimental cases the crack size can be much smaller than the sample
dimension. In such cases, detail analytical behaviors will change but physical
pictures at the scaling level stays intact.

\end{document}